\pdfoutput=1

\documentclass[11pt]{article}

\usepackage[]{ACL2023}

\usepackage{times}
\usepackage{latexsym}

\usepackage[T1]{fontenc}

\usepackage[utf8]{inputenc}

\usepackage{microtype}

\usepackage{inconsolata}

\usepackage{booktabs}
\usepackage{makecell}
\usepackage{colortbl}
\usepackage{amsmath}
\usepackage{graphicx} 
\usepackage{float} 
\usepackage{subfigure} 
\usepackage{CJKutf8}
\usepackage{multirow}

\title{Best Practices for Distilling Large Language Models into BERT for Web Search Ranking}

\author{
 \textbf{Dezhi Ye} \quad
 \textbf{Junwei Hu} \quad
 \textbf{Jiabin Fan} \quad
 \textbf{Bowen Tian} \quad 
 \textbf{Jie Liu} \quad 
 \textbf{Haijin Liang} \quad
 \textbf{Jin Ma} \quad 
\\
 Tencent
\\
\texttt{\{dezhiye, keewayhu, robertfan, lukatian,}\\
\texttt{jesangliu,  hodgeliang, daniellwang\}@tencent.com} \quad
}

\begin{document}
\maketitle
\begin{abstract}
Recent studies have highlighted the significant potential of Large Language Models (LLMs) as zero-shot relevance rankers. These methods predominantly utilize prompt learning to assess the relevance between queries and documents by generating a ranked list of potential documents. Despite their promise, the substantial costs associated with LLMs pose a significant challenge for their direct implementation in commercial search systems. To overcome this barrier and fully exploit the capabilities of LLMs for text ranking, we explore techniques to transfer the ranking expertise of LLMs to a more compact model similar to BERT, using a ranking loss to enable the deployment of less resource-intensive models. Specifically, we enhance the training of LLMs through Continued Pre-Training, taking the query as input and the clicked title and summary as output. We then proceed with supervised fine-tuning of the LLM using a rank loss, assigning the final token as a representative of the entire sentence. Given the inherent characteristics of autoregressive language models, only the final token </s> can encapsulate all preceding tokens. Additionally, we introduce a hybrid point-wise and margin MSE loss to transfer the ranking knowledge from LLMs to smaller models like BERT. This method creates a viable solution for environments with strict resource constraints. Both offline and online evaluations have confirmed the efficacy of our approach, and our model has been successfully integrated into a commercial web search engine as of February 2024.

\end{abstract}

\section{Introduction}
Relevance ranking is a paramount challenge in web search systems. The objective of relevance ranking is to rank candidate documents based on their pertinence to a specified inquiry. These documents are usually culled from an extensive corpus by a retrieval module. Of late, the integration of pre-trained language models (PLMs) such as BERT\cite{devlin2018bert}, along with industry giants like Google\footnote{https://blog.google/products/search/search-language-understanding-bert/}, Bing\footnote{https://azure.microsoft.com/en-us/blog/bing-delivers-its-largest-improvement-in-search-experience-using-azure-gpus/}, and Baidu\cite{zou2021pre,liu2021pre}, has been massively adopted within industry web search systems, yielding commendable results\cite{zhuang2023rankt5}. BERT models are adept at considering the entire context of a word by examining adjacent words, which is particularly beneficial for discerning the intent of search queries. The efficacy of IR dictates the system's response time to inquiries of users, which predominantly contingent on the performance of ranking model

The recent triumphs LLMs\cite{brown2020language} in natural language processing have ignited interest in their application to text ranking. Researchers have delved into prompting LLMs to undertake zero-shot unsupervised ranking employing pointwise\cite{wang2023query2doc,sachan2023questions}, pairwise\cite{sachan2022improving}, or listwise approaches\cite{sun2023chatgpt}. Although these have made notable strides, they have yet to fully harness the potential of LLMs. Moreover, there have been initiatives to train pointwise rankers in supervised settings, utilizing LLMs, as exemplified by RankLLaMA\cite{ma2023fine}. Despite the SOTA performance yielded by LLM rank models in experimental settings, their direct application in real-world search engines is impractical.

\begin{figure}[htbp]
\centering 
\includegraphics[width=0.49\textwidth]{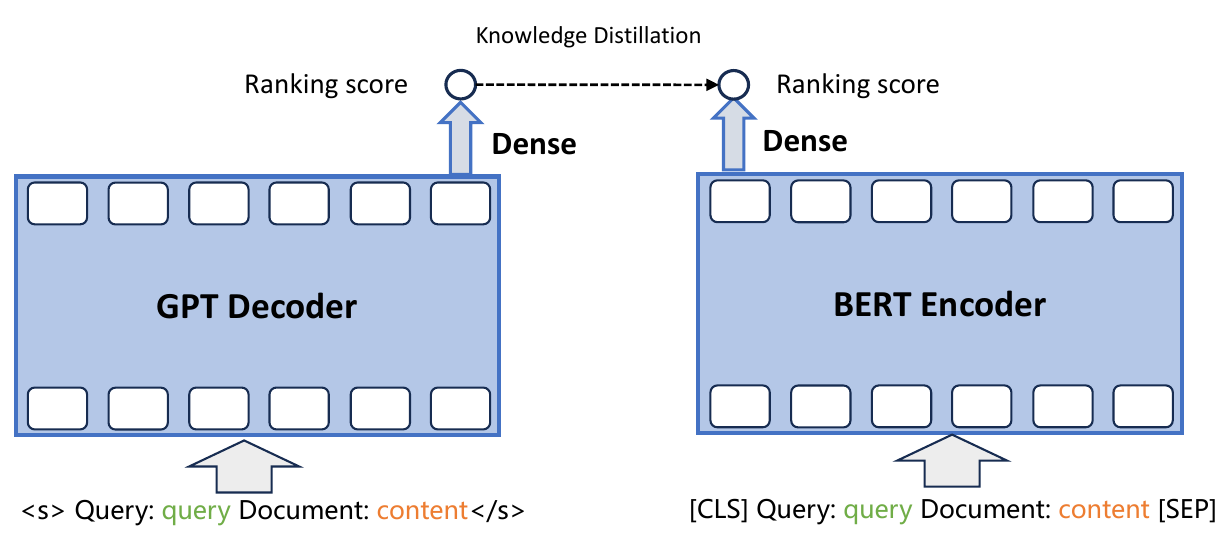}
\caption{The overview of Rank Distillation from LLM Decoder to BERT Encoder.} 
\label{Fig.intro} 
\end{figure}

To overcome the challenges of deploying LLMs online, this paper introduces a novel Rank Distillation framework (DisRanker) that combines the capabilities of LLMs with the agility of BERT. Distillation is renowned for enhancing the efficiency of various neural ranking models\cite{hofstatter2020improving}. Simultaneously, knowledge distillation facilitates the transfer of discerning skills from the teacher model to more compact models, significantly reducing computational costs during online inference. Initially, we utilize clickstream data to propagate domain knowledge through Continued Pre-Training (CPT)\cite{gupta2023continual}, using queries as inputs to generate titles and summaries that have captured user interest. In a process similar to question-answering, the LLM develops a detailed understanding of the interaction between queries and documents. We then refine the LLM using a pairwise rank loss, employing the end-of-sequence token, </s>, to represent query-document pairs. While previous research on neural rank models primarily used a bidirectional encoder-only model like BERT, interpreting the [CLS] token as a comprehensive representation of the text input, the autoregressive nature of LLMs prompts us to introduce an end-of-sequence token for the input query and document to structure the input sequence. The latent state from the final layer corresponding to this token is considered the embodiment of the query and document relationship. Consequently, we integrate a dense layer to act as a relevance adjudicator, applying pairwise rank loss to fine-tune the LLM. The deployment of LLMs for ranking tasks still faces practical challenges, particularly regarding application efficacy and output consistency. In the next phase, we employ a hybrid approach using Point-MSE\cite{qin2021improving} and Margin-MSE\cite{hofstatter2020improving} losses to distill the LLM. Point-MSE calculates the absolute difference between the LLM teacher and the BERT student, while Margin-MSE introduces a form of regularization and encourages the student model to learn the relative ranking from the teacher. This approach prevents overfitting by not requiring the student model to exactly match the teacher's scores but to maintain the order of the scores, which is essential for ranking tasks. Thus, the student model learns to emulate the teacher's ranking behavior while being more lightweight and efficient, making it better suited for deployment in resource-constrained environments.
The main contributions of this paper can be summarized as follows:
\begin{itemize}
    \item We present \textbf{DisRanker}, a novel Rank Distillation pipeline that seamlessly integrates LLM with BERT to enhance web search ranking. A comprehensive suite of offline and online evaluations substantiates the efficacy of DisRanker.
    \item We propose a domain-specific continued pre-training methods which is beneficial for enhancing the performance of LLMs on text ranking tasks. Additionally, we contribute a hybrid approach that employs Point-MSE and Margin-MSE loss to refine the distillation of LLM. 
\end{itemize}

\section{Related Work}
\subsection{LLM for Text Ranking}
Large language models have been increasingly harnessed for relevance ranking tasks in search engines\cite{sachan2023questions,muennighoff2022sgpt,wang2023query2doc}. These methodologies primarily bifurcate into two streams: one is the prompt approach\cite{qin2023large,zhuang2024setwise,ma2023zero}, and the other is the supervised fine-tuning technique\cite{zhang2023rank,ma2024fine,zhang2023rankinggpt}. In the realm of prompts, rankGPT\cite{sun2023chatgpt} has unveiled a zero-shot permutation generation method, which incites LLMs to directly generate the ranking order. Remarkably, its performance eclipses that of supervised models, particularly when utilizing GPT-4\cite{achiam2023gpt}. In the domain of supervised fine-tuning, RankLLaMA\cite{ma2023fine} injects a prompt that includes the query-document pair into the model, subsequently refining the model using a point-wise loss function. TSRankGPT \cite{zhang2023rankinggpt} advocates for a progressive, multi-stage training strategy tailored for LLMs. Indeed, while these methodologies have achieved commendable results, few have delved into how to enhance the performance of LLM models through continued pre-training, or how to effectively harness rank loss to bolster ranking capabilities.

\subsection{Knowledge Distillation for Text Ranking}

Knowledge Distillation in text ranking\cite{reddi2021rankdistil,formal2022distillation,zhuang2021ensemble,cai2022pile} indeed centers on minimizing the discrepancy between the soft target distributions of the teacher and the student\cite{tang2018ranking}. The overarching goal of distillation methods is to condense the model size and curtail the aggregate inference costs, which encompasses both memory requirements and computational overhead\cite{gao2020understanding,he2022metric}. The student model is then trained on this enriched dataset using a specialized loss function known as Margin MSE\cite{hofstatter2020improving}. Instruction Distillation\cite{sun2023instruction} proposes to distill the pairwise ranking ability of open-sourced LLMs to a simpler but more efficient pointwise ranking. This paper primarily investigates the methodology of distilling the ranking capabilities of a LLM Decoder into a BERT Encoder.

\section{Method}
\subsection{Preliminaries}
\textbf{Text Rank}. Given a query $Q$ and a candidate documents $D =\{ d_1, d_2, \cdots, d_n \} $, the task of text ranking is to compute a relevance score $S(q, d_i)$ for each document $d_i$ in $D$. The relevance labels of candidate documents with regard to the query are represented as $Y_i = (y_{i1}, . . . , y_{im})$ where $y_{ij} > 0$. We aim to optimize the ranking metrics after we sort the documents $d_i$ for each query $q_i$ based on their ranking scores. $\mathcal{L}$ is the loss function.

$$ \mathcal{L} = \sum_{q \in Q}(l(Y_{D_q},S(q,D_q))) $$

\textbf{Knowledge Distillation}. Given a large teacher model $T$ finetuned well in advance and a small student $S$, the task of knowledge distillation is to transfer $T$ to $S$ by minimizing the difference between them which can be formulated as:

$$\mathcal{L}_{KD} = \sum_{x\in \mathcal{D}} \mathbf{M}(f_{T}(x),f_{S}(x))$$

where $\mathcal{D}$ denotes the training dataset and $x$ is the input sample, $f_{T}(x),f_{S}(x)
$ represents scores of teacher and student models, and $\mathbf{M}(\cdot)$ is a loss function to measure the difference between their behaviors. 

\begin{figure*}[htbp]
\centering 
\includegraphics[width=0.99\textwidth]{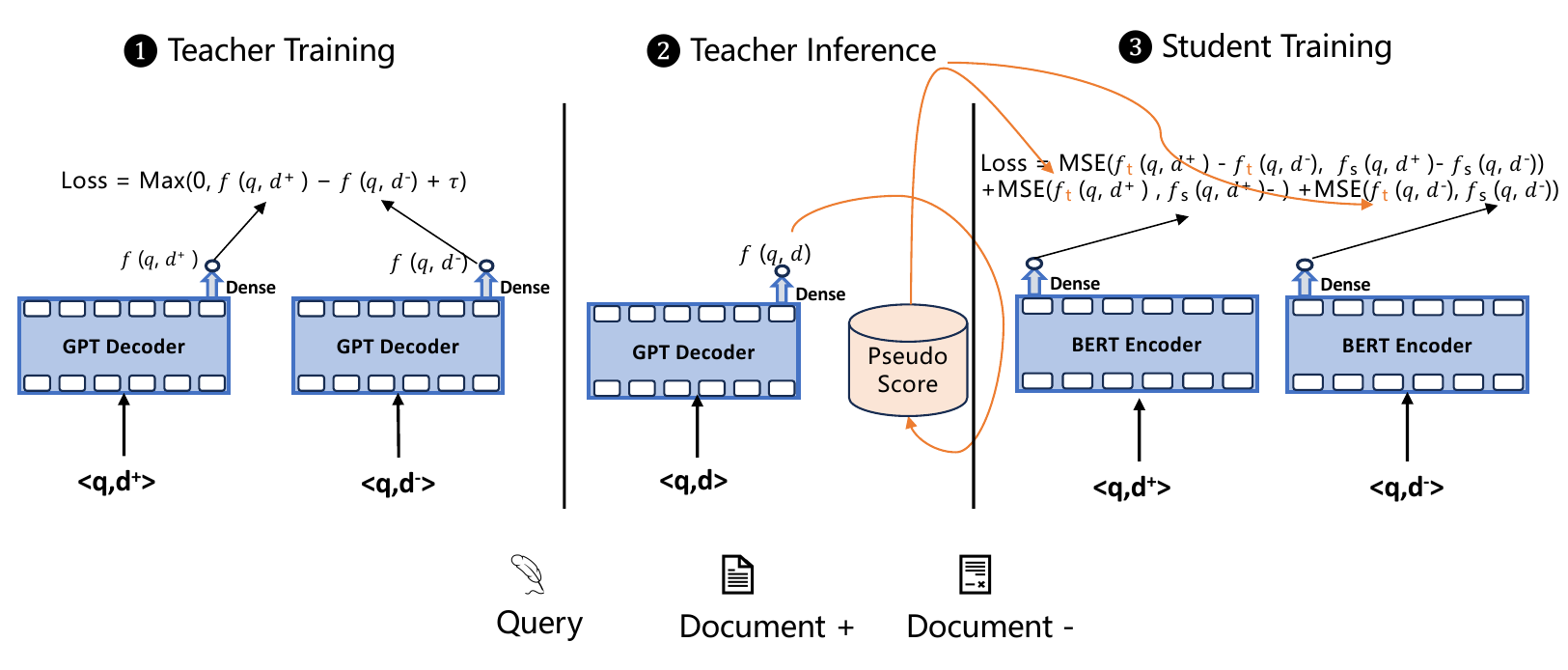}
\caption{Illustration of the knowledge distillation process: Step 1: Performing supervised fine-tuning of LLM using a ranking loss.
Step 2: Utilizing LLM to score unlabeled data. Step 3: Employing hybrid rank loss to distill the knowledge into the BERT model.} 
\label{Fig.dep} 
\end{figure*}

\subsection{Domain-Continued Pre-Training}
The pre-training task for LLMs is centered on next-token prediction, which primarily imparts general knowledge but does not inherently capture explicit signals that delineate the correlation between queries and documents. To address this, we introduce an additional phase of continue d pre-training that leverages search data to endow the model with a more refined comprehension of such relationships. Specifically, we have curated a collection of high-quality clickstream data formatted as [Query, Title, Summary]. The task of LLM is then to generate a Title and corresponding Summary based on the query, akin to a question-answering format, thereby stimulating the model's capacity to model relevance. During this stage, the tokenized raw texts of the query serve as the input.

$$\mathcal{L}_{cpt} = -\sum_j \log (T_j,S_j|P(q),q<j) $$

\begin{figure}[htbp]
\centering 
\includegraphics[width=0.4\textwidth]{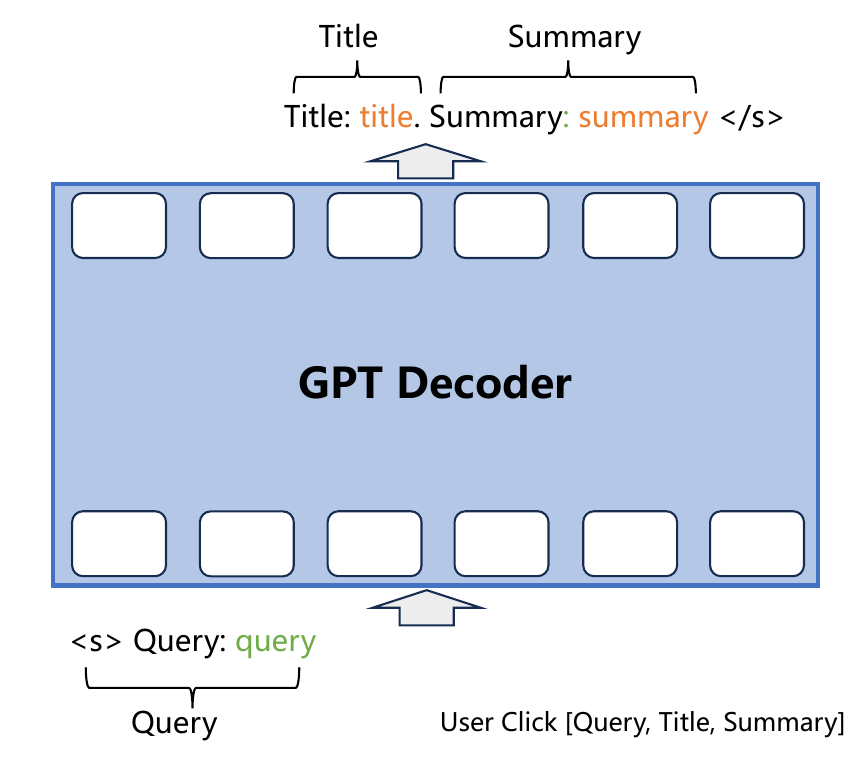}
\caption{Illustration of domain continued pre-training. The task is to generate a clicked title and summary based on a given query.} 
\label{Fig.cpt} 
\end{figure}

\subsection{Supervised Fine-Tuning}
Although LLMs have instigated a paradigm shift in natural language processing with their remarkable performance, there remains a discernible gap between the pre-training task of next-token prediction and the fine-tuning objectives. To bridge this, we append an end-of-sequence token, </s>, to the input query-document sequence to represent the entirety. Due to the autoregressive nature of the LLM model, only the final token can observe the preceding tokens, which is a distinction from BERT's approach. Concurrently, we have constructed a dense layer that maps the last layer representation of the end-of-sequence token to a scalar value.

$$ \mathrm{input = <s>query:title:summary</s>}$$
$$ f(Q,D) = Dense(Decoder(input)[-1]) $$
$$Loss = Max(0,f(q,d^+)-f(q,d^-),\mathcal{T} )$$

\subsection{Knowledge Distillation with Rank Loss}
Following the supervised fine-tuning of LLM, we conducted predictions on a large corpus of unlabeled data, then utilized the score of teacher to distill knowledge into the student models. We employed a hybrid approach of pointwise and Margin MSE loss for distillation. Considering a triplet of queries $q$, a relevant document $d^+$, and a non-relevant document $d^-$, we use the output margin of the teacher model as a label to optimize the weights of the student model as Margin MSE loss, and the output score of the teacher model as a label to optimize the weights of the student model as pointwise MSE loss.

\begin{equation*}
\begin{split}
\mathcal{L}_{Point}  & =  MSE(Sim_T(q,d^+), Sim_S(q,d^+)) \\ 
 & + MSE(Sim_T(q,d^-), Sim_S(q,d^-))\\
\end{split}
\end{equation*}

\begin{equation*}
\begin{split}
\mathcal{L}_{Margin} &= MSE(Sim_T(q,d^+) - Sim_T(q,d^-) \\ 
 &,Sim_S(q,d^+) - Sim_S(q,d^-))
\end{split}
\end{equation*}

$$\mathcal{L}_{hybrid} = \mathcal{L}_{Point} + \beta * \mathcal{L}_{Margin}$$

where $\beta$ is a scalar to balance the pointwise and Margin loss.

\begin{figure*}
	\centering
	\subfigure[LLM teacher ]{
		\begin{minipage}[b]{0.3\textwidth}
			\includegraphics[width=1\textwidth]{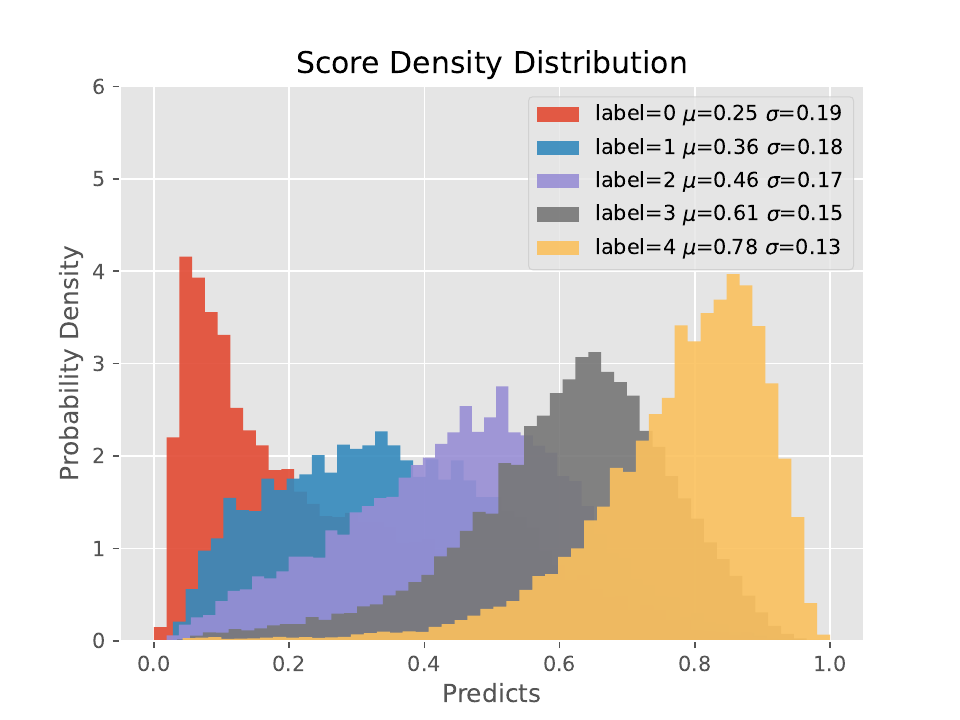}
		\end{minipage}
		\label{figdis:a}
	}
    	\subfigure[BERT-finetune ]{
    		\begin{minipage}[b]{0.3\textwidth}
   		 	\includegraphics[width=1\textwidth]{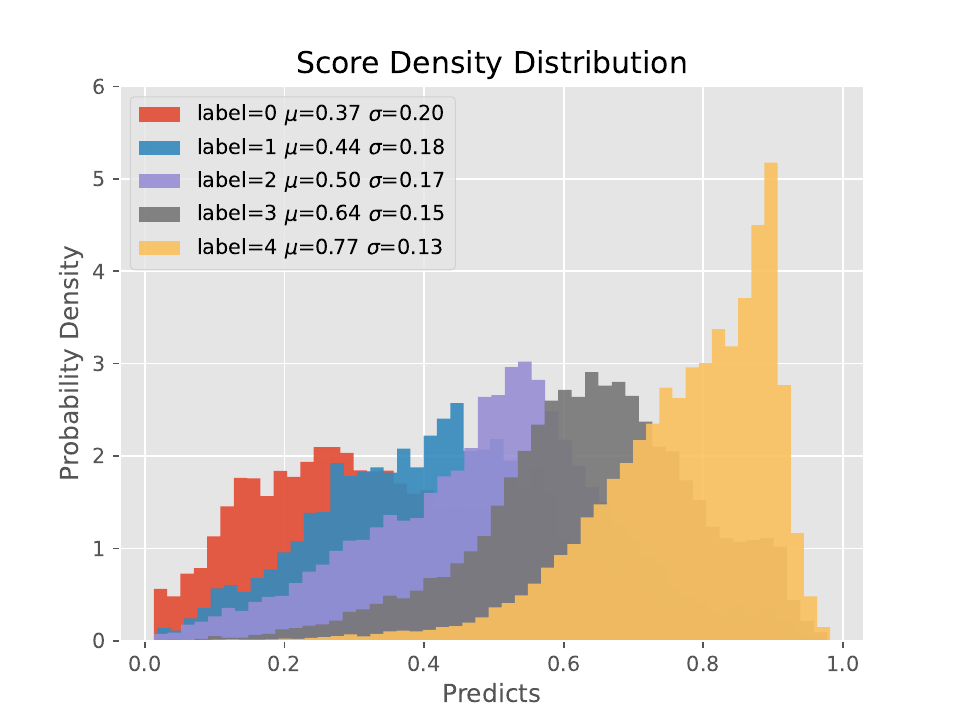}
    		\end{minipage}
		\label{figdis:b}
    	}
        \subfigure[BERT student ]{
    		\begin{minipage}[b]{0.3\textwidth}
   		 	\includegraphics[width=1\textwidth]{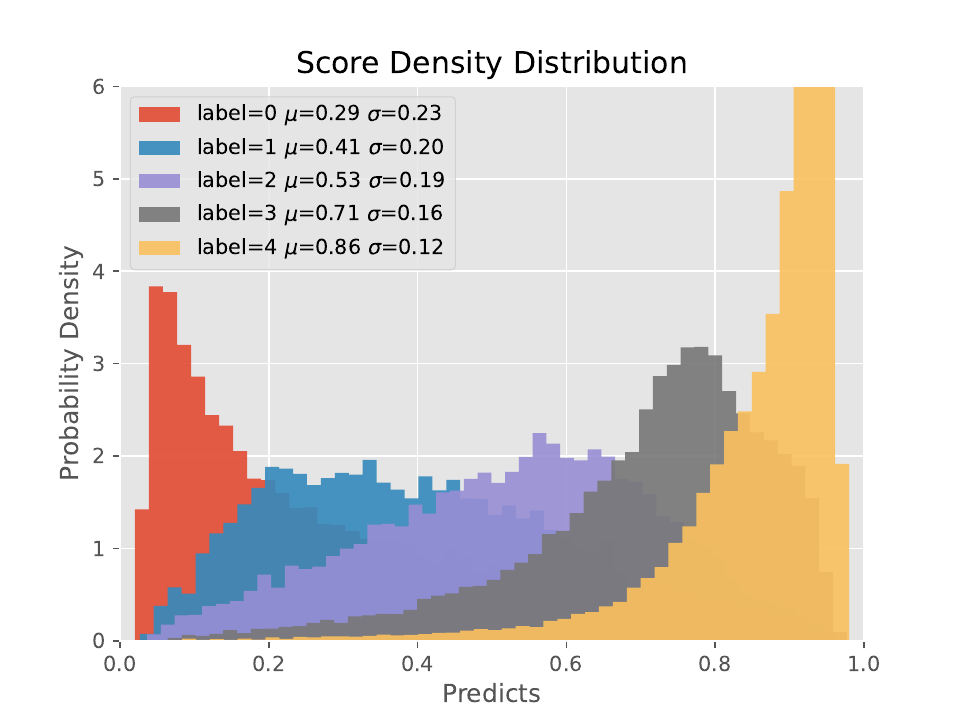}
    		\end{minipage}
		\label{figdis:c}
    	}
	\caption{The output score distribution of the LLM teacher, BERT student, and BERT-finetune models on test datasets.}
	\label{figdis}
\end{figure*}

\section{Experiments}
\subsection{Dataset and Metrics}
The datasets employed for the Continued Pre-Training(CPT), Supervised Fine-Tuning(SFT), Knowledge Distillation(KD), and test are sourced from a commercial web search engine. In the SFT phase, queries and documents extracted from search engine workflows were meticulously annotated by professionals. For the Knowledge Distillation phase, we gathered an extensive dataset from anonymous search logs, which includes 46,814,775 query-document pairs. The dataset information is summarized in Table \ref{datasets}.

\begin{table}[htbp]
\centering
\begin{tabular}{lcc}
\hline 
\textbf{Data Type} & \textbf{Queries}  & \textbf{Q-D Pairs}   \\ \hline
CPT &  10,468,974 & 59,579,125  \\ 
KD &  5,939,563 & 46,814,775 \\ \hline
SFT Data &  106,496 & 796,095 \\ 
Test Data &  13,094 & 104,960 \\ 
\hline
\end{tabular}
\caption{\label{datasets}  Statistics of datasets.}
\end{table}

In our experiments, we employed the Positive-Negative Ratio (PNR)\cite{ye2024enhancing} and Normalized Discounted Cumulative Gain (NDCG) as our principal evaluation metrics. PNR gauges the concordance between the definitive labels and the predictive scores generated by the models. NDCG, a metric ubiquitously utilized in the industry, appraises the efficacy of search engine result rankings.

\subsection{Baselines and Settings}
We conduct several comparison experiments on the following baselines. For Unsupervised LLM Rankers: Pointwise\cite{sun2023chatgpt}, Pairwise\cite{qin2023large}, Listwise\cite{ma2023zero}. For Supervised Rankers: RankLLaMA\cite{ma2023fine}, TSRankGPT\cite{zhang2023rankinggpt}, LLM2Vec\cite{behnamghader2024llm2vec}. 
In addition, we also used a fully domain-trained BERT as a baseline.
For supervised rankers, we adopt LLaMA 2\cite{jiang2023mistral} 7B as base model. For unsupervised LLM rankers, we use GPT4. In the distillation experiment, we employed LLM\footnote{https://huggingface.co/mistralai/Mistral-7B-v0.1} as the teacher model, while BERT-6L\footnote{https://huggingface.co/google-bert/bert-base-multilingual-cased} was utilized for the student model. For hyperparameter, we set $\mathcal{T}$ = 0.1, $\beta$ =0.4.

\subsection{Offline Results}

\begin{table}[htbp]
\centering
\begin{tabular}{lcc}
\hline 
\textbf{Model} & \textbf{PNR}  & \textbf{nDCG@5}   \\ \hline
BERT-Base &  3.252 & 0.8336 \\ 
BERT-large  &  3.426 & 0.8412  \\\hline
GPT4-Pointwise &  3.01 & 0.8251  \\
GPT4-Pairwise &  3.14 & 0.8273  \\
GPT4-Listwise &  3.19 & 0.8299  \\\hline
TSRankGPT &  3.475 & 0.8505 \\ 
RankLLaMA &  3.514 & 0.8611 \\
LLM2Vec &  3.496 & 0.8604 \\
\hline
DisRanker-Teacher &  3.546 & 0.8709 \\ 
\quad with CPT &  \textbf{3.643} & \textbf{0.8793}  \\ 
\hline
\end{tabular}
\caption{\label{teacher}  Offline comparison of LLM ranker performance on test sets.}
\end{table}

Unsupervised LLM rankers generally do not outperform BERT models that have been comprehensively trained within a specific domain. When compared with zero-shot methods, BERT-Base gets 3.252 on PNR but unsupervised pointwise zero-shot ranker gets 3.01. Listwise ranker achieved better results, but still can't beat BERT-Base.

Fine-tuning LLMs with rank loss significantly enhances their ranking capabilities. When compared with zero-shot methods, the NDCG@5 of RankLLaMA improve from 0.8251 to 0.8505. RankLLaMA and DisRanker are better than TSRankGPT, which underscores the effectiveness of selecting the </s> token as the representation of the query-document pair. Compared to RankLLaMA, PNR of DisRanker has increased from 3.514 to 3.546, indicating that LLM can benefit from rank loss.

Continued Pre-Training (CPT) further benefits LLM for web search ranking. The PNR improved from 3.546 to 3.643, and the NDCG increased from 0.8709 to 0.8793, which indicates that Continued Pre-Training with large-scale behavioral data substantially improves the performance of the ranking model, likely by aligning the model more closely with the specific domain and user interaction patterns.

\begin{table}[htbp]
\centering
\begin{tabular}{lcc}
\hline 
\textbf{Distill Strategy} & \textbf{PNR}  & \textbf{nDCG@5}   \\ \hline
BERT-large distill  &  3.352 &  0.8367  \\ \hline
Instruction distillation &  3.538 & 0.8464 \\ 
Point-wise &  3.534 & 0.8496 \\ 
Margin-wise &  3.554 & 0.8460 \\ 
Hybrid-loss &  \textbf{3.593} & \textbf{0.8536} \\ 
\hline
\end{tabular}
\caption{\label{studnet} Rank distill loss function ablation results on test sets. The student is a 6-layer BERT.}
\end{table}

The hybrid distillation loss enable LLM to achieve better results. Compared to using only point-wise or only margin loss, there is an improvement of 1.6\% and 1.1\% on PNR, respectively. This suggests that both the absolute scores from the teacher model and the pairwise differences provide distinct and valuable information to the student models. Furthermore, the margin MSE loss appears to be particularly effective on PNR while less help for nDCG, which may be due to its focus on the relative ranking of documents rather than absolute score predictions. Instruction distillation\cite{sun2023instruction} also achieved comparable results.


\subsection{Online A/B Test}
The online A/B test results for DisRanker, as shown in Table \ref{Online}, are quite promising. Our online baseline is a 6-layer BERT obtained by distilling a 24-layer BERT. Deploying DisRanker to the live search system and comparing it with the baseline model over the course of one week has yielded the following statistically significant improvements: PageCTR has increased by 0.47\%. The average post-click dwell time, which suggests how long users stay on the page after clicking a search result, has gone up by 1.2\%. UserCTR has increased by 0.58\%. 

In addition to these user action metrics, expert assessments were also conducted. Expert manual evaluations of 200 random queries revealed a distribution of Good vs. Same vs. Bad at 54:116:30. This expert feedback is crucial as it provides a more nuanced understanding of where the model excels and where it may require further refinement.

\begin{table}[htbp]
\centering
\begin{tabular}{lcc}
\hline 
\textbf{Metric} & \textbf{$\triangle$ Gain}& \textbf{P value} \\ \hline
PageCTR &  +0.47\% $\uparrow$ & 0.025  \\ 
UserCTR &  +0.58\% $\uparrow$ & 0.018  \\ 
Change Query Ratio &  -0.38\% $\downarrow$ & 0.026  \\ 
Dwell time &  +1.2\% $\uparrow$ &	0.016  \\ 
$\triangle_{\mathrm{GSB}}$ &  +12\% $\uparrow$& 0.002 \\ 
\hline
\end{tabular}
\caption{\label{Online} Online A/B test results of DisRanker. The $\mathbf{p}$-value is less than 0.05}
\end{table}

\subsection{Runtime Operational Improvement}
To provide a description of runtime operational improvement, we conduct an experiment comparing the LLM Teacher and the BERT student regarding throughput and cost savings. Our experiment was conducted on Nvidia A10, with the batch size set to 48. The data in Table \ref{latency} show that the LLM model consumes a considerable amount of time, which is intolerable for time-sensitive web search engines. Through distillation, we are able to conveniently transfer the capabilities of LLM to BERT, while ensuring that there is no increase in online latency.

\begin{table}[htbp]
\centering
\begin{tabular}{lcc}
\hline 
\textbf{Models} & \textbf{Params}  & \textbf{Latency}   \\ \hline
LLM  &  7B &  $\approx$100ms  \\ \hline
BERT-12 & 0.2B & $\approx$20ms \\ 
BERT-6 & 0.1B & $\approx$10ms \\ 
\hline
\end{tabular}
\caption{\label{latency} Latency between LLM teacher and BERT student. The max sequence length is set to 256.}
\end{table}

\subsection{Score Distribution Analysis}
To better understand the hybrid distillation loss, we analyze the output score distribution of the LLM teacher, BERT student, and BERT-finetuned models in Figure \ref{Online}. We observe that the score patterns between the LLM decoder and the BERT encoder models are distinct, especially at the lower and higher ends of the scoring spectrum. This discrepancy may stem from the difference in model parameter sizes, with the LLM model exhibiting higher confidence levels compared to the BERT model \cite{xiong2023can}. Employing only point-wise soft labels could potentially lead to overfitting in student models, as they might learn to replicate the teacher's output too closely without generalizing effectively. On the other hand, the margin loss introduces a form of regularization. It not only encourages the student model to learn the relative ranking from the teacher but also maintains a margin between the scores of different classes or examples.

\section{Conclusion}
In this paper, we introduce DisRanker, an innovative distillation ranking pipeline designed to harness the capabilities of LLMs for BERT. To bridge the gap between pre-training for next-token prediction and downstream relevance ranking, we initially engage in domain-specific Continued Pre-Training, using the query as input and the relevant document as output. Subsequently, we conduct supervised fine-tuning of the LLM using a ranking loss, employing the end-of-sequence token, </s>, to represent the query and document sequence. Finally, we adopt a hybrid approach of point-wise and margin MSE as our knowledge distillation loss to accommodate the diverse score output distributions. Both offline and online experiments have demonstrated that DisRanker can significantly enhance the effectiveness and overall utility of the search engine.

\section*{Ethics Statement}
The primary objective of this paper is to explore the transfer of LLM model's ranking capability to a smaller BERT model, aiming to enhance the search service provided to users. During the model training process, we have anonymized the data to ensure the protection of user privacy, without collecting any personally identifiable information.

\bibliography{anthology,custom}

\begin{thebibliography}{33}
\expandafter\ifx\csname natexlab\endcsname\relax\def\natexlab#1{#1}\fi

\bibitem[{Achiam et~al.(2023)Achiam, Adler, Agarwal, Ahmad, Akkaya, Aleman, Almeida, Altenschmidt, Altman, Anadkat et~al.}]{achiam2023gpt}
Josh Achiam, Steven Adler, Sandhini Agarwal, Lama Ahmad, Ilge Akkaya, Florencia~Leoni Aleman, Diogo Almeida, Janko Altenschmidt, Sam Altman, Shyamal Anadkat, et~al. 2023.
\newblock Gpt-4 technical report.
\newblock \emph{arXiv preprint arXiv:2303.08774}.

\bibitem[{BehnamGhader et~al.(2024)BehnamGhader, Adlakha, Mosbach, Bahdanau, Chapados, and Reddy}]{behnamghader2024llm2vec}
Parishad BehnamGhader, Vaibhav Adlakha, Marius Mosbach, Dzmitry Bahdanau, Nicolas Chapados, and Siva Reddy. 2024.
\newblock Llm2vec: Large language models are secretly powerful text encoders.
\newblock \emph{arXiv preprint arXiv:2404.05961}.

\bibitem[{Brown et~al.(2020)Brown, Mann, Ryder, Subbiah, Kaplan, Dhariwal, Neelakantan, Shyam, Sastry, Askell et~al.}]{brown2020language}
Tom Brown, Benjamin Mann, Nick Ryder, Melanie Subbiah, Jared~D Kaplan, Prafulla Dhariwal, Arvind Neelakantan, Pranav Shyam, Girish Sastry, Amanda Askell, et~al. 2020.
\newblock Language models are few-shot learners.
\newblock \emph{Advances in neural information processing systems}, 33:1877--1901.

\bibitem[{Cai et~al.(2022)Cai, Zhang, Ma, Fan, Shi, Wu, Cheng, Gu, and Yin}]{cai2022pile}
Lianshang Cai, Linhao Zhang, Dehong Ma, Jun Fan, Daiting Shi, Yi~Wu, Zhicong Cheng, Simiu Gu, and Dawei Yin. 2022.
\newblock Pile: Pairwise iterative logits ensemble for multi-teacher labeled distillation.
\newblock In \emph{Proceedings of the 2022 Conference on Empirical Methods in Natural Language Processing: Industry Track}, pages 587--595.

\bibitem[{Devlin et~al.(2018)Devlin, Chang, Lee, and Toutanova}]{devlin2018bert}
Jacob Devlin, Ming-Wei Chang, Kenton Lee, and Kristina Toutanova. 2018.
\newblock Bert: Pre-training of deep bidirectional transformers for language understanding.
\newblock \emph{arXiv preprint arXiv:1810.04805}.

\bibitem[{Formal et~al.(2022)Formal, Lassance, Piwowarski, and Clinchant}]{formal2022distillation}
Thibault Formal, Carlos Lassance, Benjamin Piwowarski, and St{\'e}phane Clinchant. 2022.
\newblock From distillation to hard negative sampling: Making sparse neural ir models more effective.
\newblock In \emph{Proceedings of the 45th International ACM SIGIR Conference on Research and Development in Information Retrieval}, pages 2353--2359.

\bibitem[{Gao et~al.(2020)Gao, Dai, and Callan}]{gao2020understanding}
Luyu Gao, Zhuyun Dai, and Jamie Callan. 2020.
\newblock Understanding bert rankers under distillation.
\newblock In \emph{Proceedings of the 2020 ACM SIGIR on International Conference on Theory of Information Retrieval}, pages 149--152.

\bibitem[{Gupta et~al.(2023)Gupta, Th{\'e}rien, Ibrahim, Richter, Anthony, Belilovsky, Rish, and Lesort}]{gupta2023continual}
Kshitij Gupta, Benjamin Th{\'e}rien, Adam Ibrahim, Mats~L Richter, Quentin Anthony, Eugene Belilovsky, Irina Rish, and Timoth{\'e}e Lesort. 2023.
\newblock Continual pre-training of large language models: How to (re) warm your model?
\newblock \emph{arXiv preprint arXiv:2308.04014}.

\bibitem[{He et~al.(2022)He, Gong, Jin, Qi, Zhang, Jiao, Zhou, Cheng, Yiu, and Duan}]{he2022metric}
Xingwei He, Yeyun Gong, A-Long Jin, Weizhen Qi, Hang Zhang, Jian Jiao, Bartuer Zhou, Biao Cheng, Sm~Yiu, and Nan Duan. 2022.
\newblock Metric-guided distillation: Distilling knowledge from the metric to ranker and retriever for generative commonsense reasoning.
\newblock In \emph{Proceedings of the 2022 Conference on Empirical Methods in Natural Language Processing}, pages 839--852.

\bibitem[{Hofst{\"a}tter et~al.(2020)Hofst{\"a}tter, Althammer, Schr{\"o}der, Sertkan, and Hanbury}]{hofstatter2020improving}
Sebastian Hofst{\"a}tter, Sophia Althammer, Michael Schr{\"o}der, Mete Sertkan, and Allan Hanbury. 2020.
\newblock Improving efficient neural ranking models with cross-architecture knowledge distillation.
\newblock \emph{arXiv preprint arXiv:2010.02666}.

\bibitem[{Jiang et~al.(2023)Jiang, Sablayrolles, Mensch, Bamford, Chaplot, Casas, Bressand, Lengyel, Lample, Saulnier et~al.}]{jiang2023mistral}
Albert~Q Jiang, Alexandre Sablayrolles, Arthur Mensch, Chris Bamford, Devendra~Singh Chaplot, Diego de~las Casas, Florian Bressand, Gianna Lengyel, Guillaume Lample, Lucile Saulnier, et~al. 2023.
\newblock Mistral 7b.
\newblock \emph{arXiv preprint arXiv:2310.06825}.

\bibitem[{Liu et~al.(2021)Liu, Lu, Cheng, Shi, Wang, Cheng, and Yin}]{liu2021pre}
Yiding Liu, Weixue Lu, Suqi Cheng, Daiting Shi, Shuaiqiang Wang, Zhicong Cheng, and Dawei Yin. 2021.
\newblock Pre-trained language model for web-scale retrieval in baidu search.
\newblock In \emph{Proceedings of the 27th ACM SIGKDD Conference on Knowledge Discovery \& Data Mining}, pages 3365--3375.

\bibitem[{Ma et~al.(2023{\natexlab{a}})Ma, Wang, Yang, Wei, and Lin}]{ma2023fine}
Xueguang Ma, Liang Wang, Nan Yang, Furu Wei, and Jimmy Lin. 2023{\natexlab{a}}.
\newblock Fine-tuning llama for multi-stage text retrieval.
\newblock \emph{arXiv preprint arXiv:2310.08319}.

\bibitem[{Ma et~al.(2024)Ma, Wang, Yang, Wei, and Lin}]{ma2024fine}
Xueguang Ma, Liang Wang, Nan Yang, Furu Wei, and Jimmy Lin. 2024.
\newblock Fine-tuning llama for multi-stage text retrieval.
\newblock In \emph{Proceedings of the 47th International ACM SIGIR Conference on Research and Development in Information Retrieval}, pages 2421--2425.

\bibitem[{Ma et~al.(2023{\natexlab{b}})Ma, Zhang, Pradeep, and Lin}]{ma2023zero}
Xueguang Ma, Xinyu Zhang, Ronak Pradeep, and Jimmy Lin. 2023{\natexlab{b}}.
\newblock Zero-shot listwise document reranking with a large language model.
\newblock \emph{arXiv preprint arXiv:2305.02156}.

\bibitem[{Muennighoff(2022)}]{muennighoff2022sgpt}
Niklas Muennighoff. 2022.
\newblock Sgpt: Gpt sentence embeddings for semantic search.
\newblock \emph{arXiv preprint arXiv:2202.08904}.

\bibitem[{Qin et~al.(2023)Qin, Jagerman, Hui, Zhuang, Wu, Shen, Liu, Liu, Metzler, Wang et~al.}]{qin2023large}
Zhen Qin, Rolf Jagerman, Kai Hui, Honglei Zhuang, Junru Wu, Jiaming Shen, Tianqi Liu, Jialu Liu, Donald Metzler, Xuanhui Wang, et~al. 2023.
\newblock Large language models are effective text rankers with pairwise ranking prompting.
\newblock \emph{arXiv preprint arXiv:2306.17563}.

\bibitem[{Qin et~al.(2021)Qin, Yan, Tay, Zhuang, Wang, Bendersky, and Najork}]{qin2021improving}
Zhen Qin, Le~Yan, Yi~Tay, Honglei Zhuang, Xuanhui Wang, Michael Bendersky, and Marc Najork. 2021.
\newblock Improving neural ranking via lossless knowledge distillation.
\newblock \emph{arXiv preprint arXiv:2109.15285}.

\bibitem[{Reddi et~al.(2021)Reddi, Pasumarthi, Menon, Rawat, Yu, Kim, Veit, and Kumar}]{reddi2021rankdistil}
Sashank Reddi, Rama~Kumar Pasumarthi, Aditya Menon, Ankit~Singh Rawat, Felix Yu, Seungyeon Kim, Andreas Veit, and Sanjiv Kumar. 2021.
\newblock Rankdistil: Knowledge distillation for ranking.
\newblock In \emph{International Conference on Artificial Intelligence and Statistics}, pages 2368--2376. PMLR.

\bibitem[{Sachan et~al.(2022)Sachan, Lewis, Joshi, Aghajanyan, Yih, Pineau, and Zettlemoyer}]{sachan2022improving}
Devendra Sachan, Mike Lewis, Mandar Joshi, Armen Aghajanyan, Wen-tau Yih, Joelle Pineau, and Luke Zettlemoyer. 2022.
\newblock Improving passage retrieval with zero-shot question generation.
\newblock In \emph{Proceedings of the 2022 Conference on Empirical Methods in Natural Language Processing}, pages 3781--3797.

\bibitem[{Sachan et~al.(2023)Sachan, Lewis, Yogatama, Zettlemoyer, Pineau, and Zaheer}]{sachan2023questions}
Devendra~Singh Sachan, Mike Lewis, Dani Yogatama, Luke Zettlemoyer, Joelle Pineau, and Manzil Zaheer. 2023.
\newblock Questions are all you need to train a dense passage retriever.
\newblock \emph{Transactions of the Association for Computational Linguistics}, 11:600--616.

\bibitem[{Sun et~al.(2023{\natexlab{a}})Sun, Chen, Ma, Yan, Wang, Ren, Chen, Yin, and Ren}]{sun2023instruction}
Weiwei Sun, Zheng Chen, Xinyu Ma, Lingyong Yan, Shuaiqiang Wang, Pengjie Ren, Zhumin Chen, Dawei Yin, and Zhaochun Ren. 2023{\natexlab{a}}.
\newblock Instruction distillation makes large language models efficient zero-shot rankers.
\newblock \emph{arXiv preprint arXiv:2311.01555}.

\bibitem[{Sun et~al.(2023{\natexlab{b}})Sun, Yan, Ma, Ren, Yin, and Ren}]{sun2023chatgpt}
Weiwei Sun, Lingyong Yan, Xinyu Ma, Pengjie Ren, Dawei Yin, and Zhaochun Ren. 2023{\natexlab{b}}.
\newblock Is chatgpt good at search? investigating large language models as re-ranking agent.
\newblock \emph{arXiv preprint arXiv:2304.09542}.

\bibitem[{Tang and Wang(2018)}]{tang2018ranking}
Jiaxi Tang and Ke~Wang. 2018.
\newblock Ranking distillation: Learning compact ranking models with high performance for recommender system.
\newblock In \emph{Proceedings of the 24th ACM SIGKDD international conference on knowledge discovery \& data mining}, pages 2289--2298.

\bibitem[{Wang et~al.(2023)Wang, Yang, and Wei}]{wang2023query2doc}
Liang Wang, Nan Yang, and Furu Wei. 2023.
\newblock Query2doc: Query expansion with large language models.
\newblock \emph{arXiv preprint arXiv:2303.07678}.

\bibitem[{Xiong et~al.(2023)Xiong, Hu, Lu, Li, Fu, He, and Hooi}]{xiong2023can}
Miao Xiong, Zhiyuan Hu, Xinyang Lu, Yifei Li, Jie Fu, Junxian He, and Bryan Hooi. 2023.
\newblock Can llms express their uncertainty? an empirical evaluation of confidence elicitation in llms.
\newblock \emph{arXiv preprint arXiv:2306.13063}.

\bibitem[{Ye et~al.(2024)Ye, Liu, Fan, Tian, Zhou, Chen, and Ma}]{ye2024enhancing}
Dezhi Ye, Jie Liu, Jiabin Fan, Bowen Tian, Tianhua Zhou, Xiang Chen, and Jin Ma. 2024.
\newblock Enhancing asymmetric web search through question-answer generation and ranking.
\newblock In \emph{Proceedings of the 30th ACM SIGKDD Conference on Knowledge Discovery and Data Mining}, pages 6127--6136.

\bibitem[{Zhang et~al.(2023{\natexlab{a}})Zhang, Zhang, Long, Xie, Zhang, and Zhang}]{zhang2023rankinggpt}
Longhui Zhang, Yanzhao Zhang, Dingkun Long, Pengjun Xie, Meishan Zhang, and Min Zhang. 2023{\natexlab{a}}.
\newblock Rankinggpt: Empowering large language models in text ranking with progressive enhancement.
\newblock \emph{arXiv preprint arXiv:2311.16720}.

\bibitem[{Zhang et~al.(2023{\natexlab{b}})Zhang, Hofst{\"a}tter, Lewis, Tang, and Lin}]{zhang2023rank}
Xinyu Zhang, Sebastian Hofst{\"a}tter, Patrick Lewis, Raphael Tang, and Jimmy Lin. 2023{\natexlab{b}}.
\newblock Rank-without-gpt: Building gpt-independent listwise rerankers on open-source large language models.
\newblock \emph{arXiv preprint arXiv:2312.02969}.

\bibitem[{Zhuang et~al.(2021)Zhuang, Qin, Han, Wang, Bendersky, and Najork}]{zhuang2021ensemble}
Honglei Zhuang, Zhen Qin, Shuguang Han, Xuanhui Wang, Michael Bendersky, and Marc Najork. 2021.
\newblock Ensemble distillation for bert-based ranking models.
\newblock In \emph{Proceedings of the 2021 ACM SIGIR International Conference on Theory of Information Retrieval}, pages 131--136.

\bibitem[{Zhuang et~al.(2023)Zhuang, Qin, Jagerman, Hui, Ma, Lu, Ni, Wang, and Bendersky}]{zhuang2023rankt5}
Honglei Zhuang, Zhen Qin, Rolf Jagerman, Kai Hui, Ji~Ma, Jing Lu, Jianmo Ni, Xuanhui Wang, and Michael Bendersky. 2023.
\newblock Rankt5: Fine-tuning t5 for text ranking with ranking losses.
\newblock In \emph{Proceedings of the 46th International ACM SIGIR Conference on Research and Development in Information Retrieval}, pages 2308--2313.

\bibitem[{Zhuang et~al.(2024)Zhuang, Zhuang, Koopman, and Zuccon}]{zhuang2024setwise}
Shengyao Zhuang, Honglei Zhuang, Bevan Koopman, and Guido Zuccon. 2024.
\newblock A setwise approach for effective and highly efficient zero-shot ranking with large language models.
\newblock In \emph{Proceedings of the 47th International ACM SIGIR Conference on Research and Development in Information Retrieval}, pages 38--47.

\bibitem[{Zou et~al.(2021)Zou, Zhang, Cai, Ma, Cheng, Wang, Shi, Cheng, and Yin}]{zou2021pre}
Lixin Zou, Shengqiang Zhang, Hengyi Cai, Dehong Ma, Suqi Cheng, Shuaiqiang Wang, Daiting Shi, Zhicong Cheng, and Dawei Yin. 2021.
\newblock Pre-trained language model based ranking in baidu search.
\newblock In \emph{Proceedings of the 27th ACM SIGKDD Conference on Knowledge Discovery \& Data Mining}, pages 4014--4022.

\end{thebibliography}
\bibliographystyle{acl_natbib}

\appendix

\section{Appendix}
\label{sec:appendix}

\subsection{Industry Datasets}
The industry datasets are collected from the search pipelines and manually labeled on the crowdsourcing platform, where a group of hired annotators assigned an integer label range from 0 to 4 to each query document pair, representing their relevance as \{bad, fair, good, excellent, perfect\}.

\subsection{Evaluation Metrics}
$$ \triangle_{\mathrm{GSB}} = \frac{\#Good - \#Bad}{\#Good + \#Same + \#Bad} $$
where $\#Good$ (or $\#Bad$) donates the number of queries that the new (or base) model provides better ranking results. and $\#Same$ for the number of results that having same quality.
$$PNR = \frac{\sum_{i,j\in [1,N]} \mathbf{I}\{y_i>y_j\} \mathbf{I} \{f(q,d_i)>f(q,d_j)\} }{\sum_{i,j\in [1,N]} \mathbf{I}\{y_i>y_j\} \mathbf{I} \{f(q,d_i)<f(q,d_j)\} } $$
where the indicator function $\mathbf{I}( y_i > y_j)$ takes the value 1 if $y_i > y_j$ and 0 otherwise.

\end{document}